\documentclass[epj]{svjour}
\usepackage{graphics}
\begin{document}
\title{Proton-deuteron radiative capture cross sections at intermediate energies}
\subtitle{}

\author{A.A.~Mehmandoost-Khajeh-Dad\inst{1}\and 
        M.~Mahjour-Shafiei\inst{2}\and 
        H.R.~Amir-Ahmadi\inst{3}\and
        J.C.S.~Bacelar\inst{3}\and
        A.M.~van~den~Berg\inst{3}\and
        R.~Castelijns\inst{3}\and
        E.D.~van~Garderen\inst{3}\and 
        N.~Kalantar-Nayestanaki\inst{3}\and
        M.~Ki\v{s}\inst{4}\and
        H.~L\"ohner\inst{3}\and
        J.G.~Messchendorp \inst{3}\and
        H.J.~W\"ortche\inst{3}
}                     

\institute{Department of Physics, University of Sistan and Baluchestan, P.O. Box 98155/987 Zahedan, Iran\and 
           Department of Physics, University of Tehran, P.O. Box 14395/547, North Kargar Avenue, Tehran Iran\and 
           Kernfysisch Versneller Instituut (KVI), University of Groningen, Zernikelaan 25, 9747 AA Groningen, The Netherlands\and
           Rudjer Boskovic Institute, Zagreb, Croatia
}
\date{Received: date / Revised version: date}
%
\abstract{
Differential cross sections of the reaction $p(d,^3{\rm He})\gamma$ have been measured at deuteron 
laboratory energies of 110, 133 and 180 MeV. The data were obtained with a coincidence setup measuring 
both the outgoing $^3$He and the photon. The data are compared with modern calculations including 
all possible meson-exchange currents and two- and three- nucleon forces in the potential. The data 
clearly show a preference for one of the models, although the shape of the angular distribution cannot 
be reproduced by any of the presented models.
\keywords {Nuclear force--Radiative capture--Differential Cross Section--Geant3}
\PACS{
      {}{21.30.Fe; 21.45.+v; 25.10.+s} 
     } 
} 
\maketitle
\section{Introduction}
\label{intro}

Few-nucleon systems have been extensively used in the past to investigate various facets of the nuclear forces. 
Nucleon-nucleon forces (2NF) are now quite well established and modern potentials predict all possible observables 
with a reduced $\chi^2$ very close to unity~\cite{Stok1993}. Reactions which involve more than two nucleons probe 
parts of 2NF which are not directly accessible in nucleon-nucleon scattering. These reactions could involve a real 
or virtual photon in the final state such as in bremsstrahlung process or a third nucleon present in the interaction. 
Precise measurements in the past exploiting in a bremsstrahlung process showed clear disagreements between the theoretical 
predictions and the experimental data~\cite{Messchendorp99-1,Messchendorp99-2,Huisman99-1,Hoefman,messchendorp2000-1,messchendorp2000-2,Huisman2000,Huisman2001,volkerts2003,volkerts2004,shafiei2004,shafiei2009} although recent attempts in theory seem to alleviate some of the discrepancies~\cite{Nakayama2009}. 
Also the proton-deuteron scattering process at 
intermediate energies showed a need for three-nucleon forces (3NF) in the 
Hamiltonian and, even then, the present models for these forces are not 
adequate to describe all of the data~\cite{Bieber2000,Ermisch2001,Kistryn2003,Ermisch2003,Sekiguchi2004,Ermisch2005,Kistryn2005,Mardanpour2007,Amirahmadi2007,Stephan2007,Ramazani2008,Stephan2009,Mardanpour2010,Johansson98,Mes09,Glo96,Glo01,Kam01}. 
There is an important difference in the two processes mentioned. 
The bremsstrahlung process involves a photon in the 
final state, thereby emphasizing the role of Meson-Exchange Currents (MEC) while in the case of proton-deuteron 
scattering only hadrons are present in the initial and in the final states. In that sense, the off-shell behavior 
of particles could be different depending on the momentum with which the interaction is probed. The radiative capture 
process is a particular case in which large momentum mis-matches are involved making this process 
very attractive to probe the high-momentum parts of the wave functions. The two-body radiative capture has been 
extensively studied in the past and physical observables could only be described once MEC were taken into account. 

For the three-body hadronic scattering, the 2NF can be used in a Faddeev calculation which produces exact results. 
Here, the problem resides in modeling the right 3NF for this process. For the three-body bremsstrahlung process, 
no attempt has been made to implement MEC in a potential model calculation. For the proton-deuteron 
radiative capture process, both the MEC and 3NF should be taken into account. 
For intermediate photon center-of-mass energies (up to 100 MeV), the effects of the 3NF are shown to be rather 
small making this process ideal to study the MEC at the specific kinematics of the capture process. Data are rather 
scarce on this process. Experiments have been performed either with a proton or deuteron beam.  
Most of these experiments used a solid target (in the form of CH$_2$ or CD$_2$) making 
it necessary to perform a careful study of the backgrounds~\cite{Tam07,Yag03,Pic86,Pitts1988}. 

In this paper, we report on an exclusive measurement of differential cross sections of the proton-deuteron radiative 
capture process obtained at incident polarized deuteron-beam energies of 110, 133 and 180~MeV. The analyzing powers 
obtained in the same measurement have already been published~\cite{ali}. Here, the cross section data will be presented 
and compared with theoretical models. The first model developed by the Bochum-Cracow groups~\cite{Ski03,Gol00} is a 
rigorous Faddeev calculation with the Argonne V18 (AV18) 2NF as input with the addition of the Urbana-IX 3NF. 
The coupling with a photon is described via two different approaches. The first approach supplements the single-nucleon 
current operator by exchange currents which take explicitly into account $\pi$- and $\rho$-like meson-exchange 
contributions. Alternatively, the meson-exchange currents are included using the extended Siegert theorem. In this form, 
electric and magnetic multipoles are kept to very high orders for the one-body operator. As a consequence of the 
Siegert theorem, only many-body currents in the electric multipoles are accounted for. The second calculation is 
from the Hannover-Lisbon theory groups~\cite{Del04}, which describes the process using the purely nucleonic 
charge-dependent CD-Bonn potential and its coupled-channel extension CD-Bonn+$\Delta$. Within this approach, 
the $\Delta$-isobar excitation mediates an effective 3NF with prominent Fujita-Miyazawa and Illinois ring-type 
contributions. These contributions are based on the exchanges of $\pi$, $\rho$, $\omega$, and $\sigma$ mesons and 
are mutually consistent. The electromagnetic current in the Hannover-Lisbon approach has one-baryon and two-baryon 
contributions and couples to nucleonic and $\Delta$-isobar channels. Therefore, the $\Delta$-isobar generates 
consistently effective two- and three-nucleon currents in addition to a 3NF.

\begin{figure}
\resizebox{0.50\textwidth}{!}{%
 \includegraphics{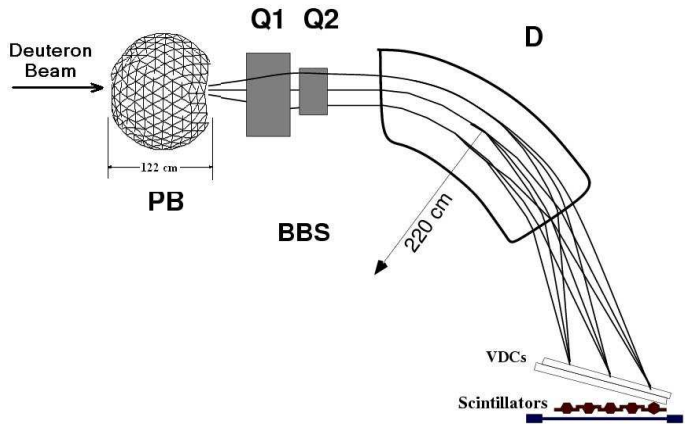}
}
\caption{Cut through the mid plane of the BBS and the corresponding detector systems. The beam first enters the 
Plastic Ball (PB) and hits the target located in the center of the PB before entering the BBS. Particles with the 
same momenta are focused to the same position in the focal plane of the magnet (where VDCs are located), 
regardless of their initial direction. The two segmented scintillators which are located after the last VDC are 
used as hardware triggers and used to identify the $^3$He.}
\label{setup}       
\end{figure}
\section{Experimental setup}
\label{Exp}
\begin{figure}
\resizebox{0.50\textwidth}{!}{%
 \includegraphics{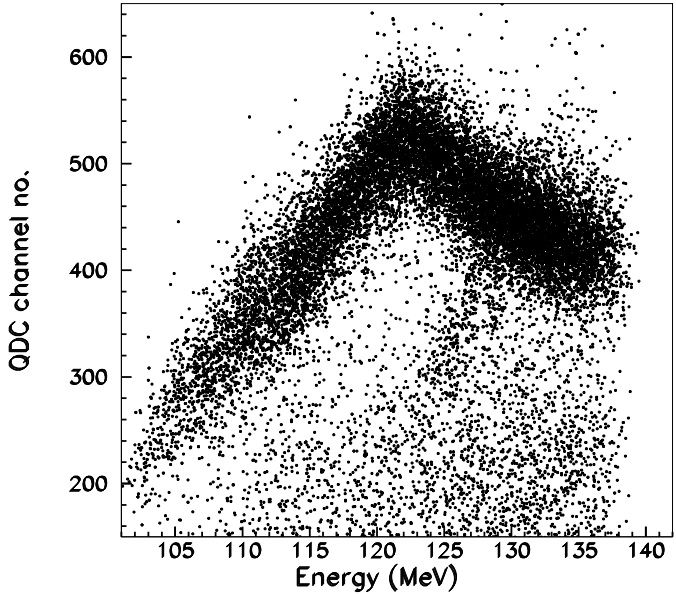}
}
\caption{The QDC channel number of the scintillators of the BBS plotted against the reconstructed energy of the $^3$He 
passing through the scintillators which determine the trigger. At least one photon is required to have been detected by the PB. 
The BBS was placed at a scattering angle of 3.5$^\circ$ and the deuteron-beam energy was 180~MeV.}
\label{scint}       
\end{figure}

\begin{figure}
\resizebox{0.50\textwidth}{!}{%
 \includegraphics{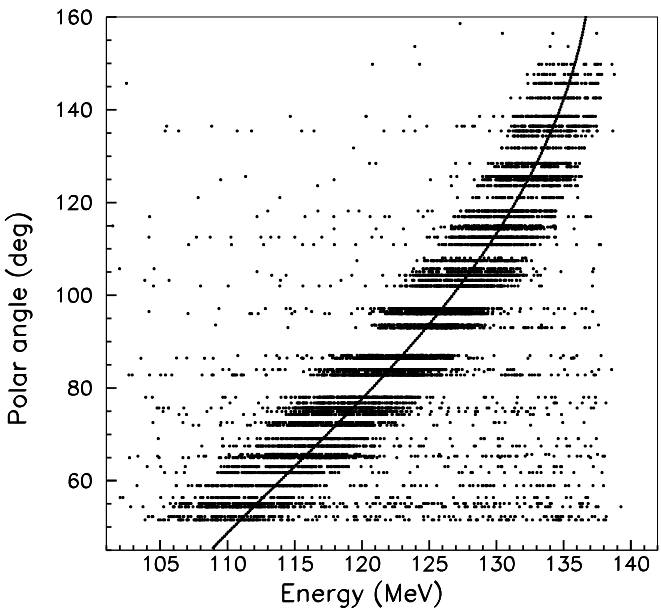}
}
\caption{The laboratory polar angle of detector modules of the PB plotted against the reconstructed energy of the $^3$He. 
The events were selected using the same conditions as used for the data shown in Fig.~\ref{scint}. The expected 
kinematical correlation of the proton-deuteron radiative capture reaction is shown by a solid line.}
\label{photon}       
\end{figure}

The experiment was performed in 2003 at the Kernfysisch Versneller Instituut (KVI) in Groningen, The Netherlands. 
A polarized deuteron beam was produced with the superconducting cyclotron, Acc{\'e}l{\'e}rateur Groningen ORsay (AGOR). 
The beam with an intensity of $\approx$0.5~nA impinged on a 58$\pm$3~mg/cm$^2$ liquid-hydrogen target \cite{Kal98} placed at 
the center of the Plastic Ball (PB)~\cite{shafiei2004,shafiei2009,Bad82,ShafieiThesis}, which was used to detect the photons coming from the radiative capture reaction. The $^3$He ions were detected with the Big-Bite Spectrometer (BBS)~\cite{Ber95}.
 Figure~\ref{setup} 
shows the experimental setup used in this experiment. The BBS is a QQD-type spectrometer and has, therefore, two 
quadrupole magnets (Q) and one dipole magnet (D) with an angular acceptance of $\approx${3.8}$^\circ$. The instrument 
is designed such that an image of a target spot is produced at the focal plane. At the position of this focal plane, 
two vertical drift chambers (VDCs) were mounted, measuring the position and the angle of the $^3$He traversing the 
focal plane. After the last VDC, two segmented scintillators are located. The event trigger is based on a coincidence 
of signals from the two scintillator planes. The response of the last scintillator is shown in Fig.~\ref{scint} 
for a beam energy of 180~MeV. This scintillator is made of NE102A with the thickness of 8~mm. The signals of the 
scintillator are integrated and digitized by charge-integrating QDCs. In this figure, the digitized information of the 
QDC, corresponding to the deposited energy in the scintillator, is plotted against the reconstructed energy of $^3$He 
from the momentum analysis of the spectrometer for the cases when at least one photon is detected by the PB. The coincidence condition reduces the peak to noise ratio to about 12/1. A clear 
band can be observed corresponding to $^3$He particles which punch through the scintillator at an energy of around 123~MeV.
Figure~\ref{photon} shows the laboratory polar angle of detector modules of the PB plotted against the reconstructed energy
of $^3$He for the same events as shown in Fig.~\ref{scint}. A clear correlation can be observed which matched perfectly the
expected kinematical correlation of the reaction of interest as shown by the solid line.

\section{Data analysis and results}
\label{Ana}
\begin{figure} 
\resizebox{0.50\textwidth}{!}{%
\includegraphics{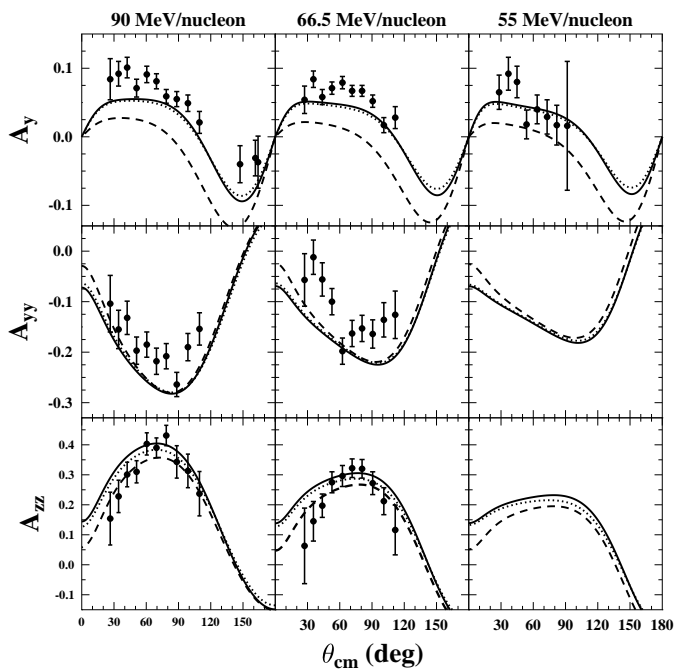}
}
\caption{Polarization data for the deuteron-proton radiative capture reaction are compared to Faddeev calculations by the Bochum-Cracow theory group~\cite{ali}. The dashed lines represent the results of the calculation using the Siegert approximation with the AV18 2NF 
as input and with the additional inclusion of the Urbana-IX 3NF. The dotted (2NF) and solid (2NF+3NF) lines are 
similar calculations for which meson-exchange currents are calculated using explicit $\pi$ and $\rho$ exchanges.}
\label{arg}       
\end{figure}

\begin{figure} 
\resizebox{0.50\textwidth}{!}{%
\includegraphics{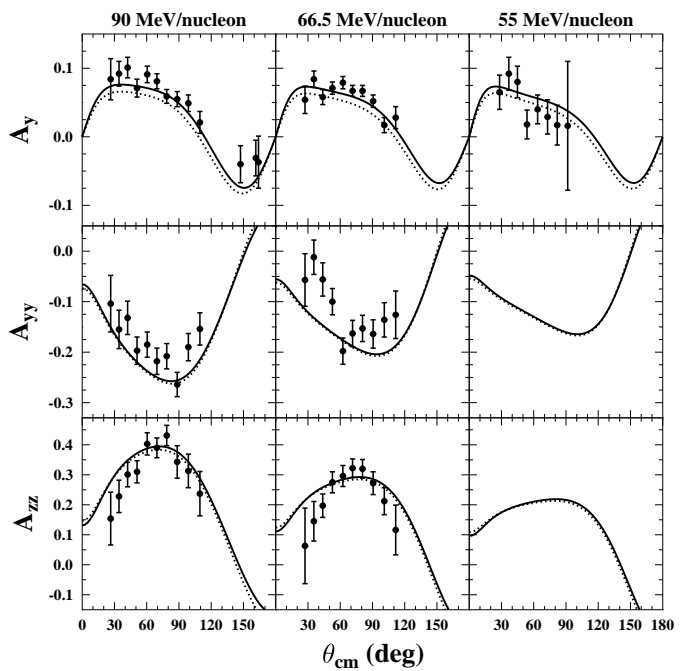}
}
\caption{Polarization data for the deuteron-proton radiative capture reaction are compared to calculations by the Hannover theory group~\cite{ali}. The solid (dotted) lines represent the results of a coupled-channel calculation based on the CD-Bonn potential with (without) an intermediate $\Delta$-isobar.}
\label{bon}       
\end{figure}

The precision of the measured absolute cross section depends strongly on the knowledge of the 
photon detection efficiency of the PB. Due to the low $Z$ of the organic 
scintillators of the Plastic Ball, some photons do not interact with the material or if they do, they leave energies 
which are below the detection threshold. The corresponding efficiency was obtained with the help of the GEANT3
transport code as explained in~\cite{ShafieiThesis} and found to be around 50\%. The exact number depends on the photon energy and 
the actual threshold for each detector which was obtained through a careful calibration of each scintillator element. 
The threshold varied between 5 to 15 MeV for all the crystals. In the simulations, care was taken to account for 
those detectors which were not operating properly during the measurement. In addition, the PB covers all 
azimuthal angles whereas the BBS, which detects the coincident $^3$He particles, has a small coverage in the 
azimuthal angle. This geometrical effect was also calculated and corrected for. 
The efficiency of the $^3$He detection and reconstruction has been unambiguously determined using the events from the 
radiative capture process. This analysis yields efficiencies ranging from 65-95\% depending on the energy of the $^3$He.  
The dominant part of the inefficiency is related to a deficiency in the reconstruction of the $^3$He momentum and is well 
understood. 
The thickness of the liquid-hydrogen target was determined from a concurrent measurement of the counts of the 
deuteron-proton elastic scattering process. The obtained experimental unnormalized cross sections were compared to the 
known cross sections of this reaction at the same energy. This comparison yielded a target thickness of 58$\pm$3~mg/cm$^2$. 
The uncertainty of the target thickness was obtained from a quadratic sum of the various individual uncertainties, 
namely 1\% statistical accuracy of the measurement of the elastic cross section, 2\% due to the uncertainty in the 
reconstruction, 4\% due to the binning in the center of mass, and 4\% due to the cross sections used in the comparison. 
As a cross check, an additional measurement of the radiative capture cross section was conducted using a solid 
CH$_2$ target with a thickness of 12.0$\pm$0.5~mg/cm$^2$. Due to large energy loss of the $^3$He in the 
liquid-hydrogen target, the experiment with the 110~MeV deuteron beam was also performed using the solid CH$_2$ target. 
The absolute cross sections of this study were found to be consistent with the measurement using the liquid-hydrogen 
target.
As mentioned earlier, polarized deuteron beams were used in these measurements making it possible to obtain 
analyzing powers in addition to cross sections. These observables suffer much less from normalization problems and 
were, therefore, produced first. The results have already been published for the vector and tensor analyzing 
powers~\cite{ali} and are shown in Figs.~\ref{arg} and ~\ref{bon}. The main goal of the present work is to present the corresponding differential cross sections.

\begin{figure*}
\resizebox{1.0\textwidth}{!}{%
 \includegraphics{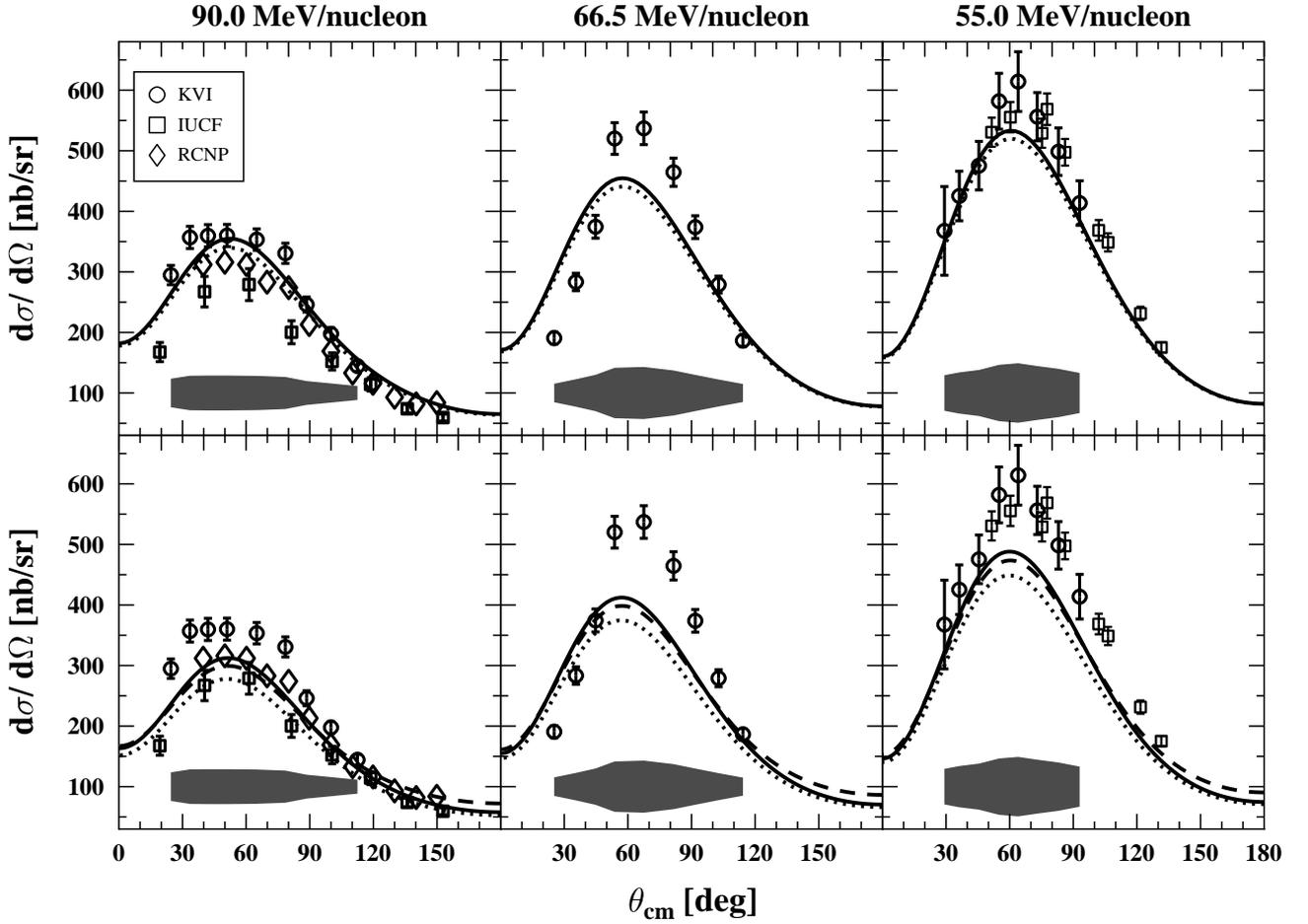}
}

\caption{Differential cross sections of the proton-deuteron radiative capture process at 180~MeV, 133~MeV and 110~MeV 
deuteron beam energies as a function of the center-of-mass angles (photon-proton angle).
 The results of the present work are plotted as open circles,
 while those of Ref.~\cite{Pic86}, with a 99.1~MeV proton beam (left panels) and those of Ref.~\cite{Pitts1988},
 with a 95~MeV deuteron beam (right panels) are represented by open squares.
 The results of Ref.~\cite{Yag03}, with a 200~MeV deuteron beam are represented by open diamonds.
 For the present work, only the statistical uncertainty is shown. The overall systematic uncertainty of $\pm$8\% is indicated as a grey band with a thickness of 2$\sigma$ in each panel. The results are compared to Faddeev calculations by the Hannover-Lisbon group in the top panels where the solid (dotted) lines represent the results of a coupled-channel calculation based on the CD-Bonn potential with (without) an intermediate $\Delta$-isobar. The bottom panels show the comparison of data with the predictions of the Bochum-Cracow theory group. 
Here, the dashed lines represent the results of the calculation using the Siegert approximation with the AV18 2NF 
as input and with the additional inclusion of the Urbana-IX 3NF. The dotted (2NF) and solid (2NF+3NF) lines are 
similar calculations for which meson-exchange currents are calculated using explicit $\pi$ and $\rho$ exchanges.}
\label{cross2}       
\end{figure*}

Our cross section data are depicted as open circles in Fig.~\ref{cross2} as a function of the center-of-mass angle of the photon-proton system. The uncertainties shown in the figure are a quadratic sum of a 5\% point-to-point (PTP) systematic uncertainty and the pure statistical uncertainty. The PTP error stems from the uncertainty in the threshold determination of the Plastic Ball (PB) detectors. These thresholds are estimated independently for each ring in the PB, and therefore vary as a function of the center-of-mass angle. The uncertainties do not include the overall uncertainties related to the target thickness ($\approx$ 6\%), the $^3$He detection ($\approx$ 2\%), and the photon detection efficiency ($\approx$ 5\%) yielding a total overall systematic uncertainty of 8\%.  
Our data are compared to results from other experiments taken at 99.1~MeV/nucleon~\cite{Pic86}, 100~MeV/nucleon~\cite{Yag03}, and 47.5~MeV/nucleon~\cite{Pitts1988}. The KVI data lie about 10\% above the other data sets after taking the energy differences into acount and ignoring some shape differences. All the data agree within one standard deviation from each other considering the systematic uncertainties and the differences in incident beam energy.

In the top panels in Fig.~\ref{cross2}, the cross section data are compared with Faddeev calculations by the 
Hannover-Lisbon group where the solid (dotted) lines represent the predictions of a coupled-channel calculation based 
on the CD-Bonn potential with (without) an intermediate $\Delta$-isobar. The $\Delta$ mediates 3NFs and generates 
effective two- and three- nucleon currents in addition to irreducible one- and two-baryon contributions.
The bottom panels show the same data compared with predictions by the Bochum-Cracow group. Here, the dotted lines are the results 
of the Faddeev calculations with the AV18 2NF used as input. The solid lines are produced with the same model but 
now including the Urbana-IX 3NF as well. For both line types, the meson exchange-currents are obtained using 
explicit $\pi$ and $\rho$ exchanges. The dashed lines represent the result of the calculation using 
the Siegert approximation with the AV18 2NF as input and with the additional inclusion of the Urbana-IX 3NF. 

The present data seem to support, in magnitude, the results of the Hannover-Lisbon calculations. The Bochum-Cracow 
results are underestimating the data. However, the angular distribution, in particular at 66.5~MeV/nucleon, differs 
from the theoretical one. The width of the peak seems to become narrower as a function of decreasing energy 
and the change is faster than observed in the theoretical predictions, in particular for 66.5~MeV/nucleon. 
We furthermore note that there are only small differences between the predictions exploiting an explicit $\pi$ and $\rho$
 exchange treatment and those which are derived from the Siegert approximation for the angular range covered by the experiment. 
Only at large backward angles one can observe significant relative differences between the predictions of the two approaches
 as shown in the bottom row in Fig.~\ref{cross2}. These differences can intuitively 
be attributed to the large magnetic contribution in the electromagnetic current at backward angles. In this case, the
Siegert approximation is known to fail. For the analyzing powers, in particular for $A_y$, as shown in Fig.~\ref{arg}, the predictions using the Siegert approximation are far from the full calculations showing that for these observables, the electric part of the MEC is not sufficient to describe the whole angular range. It is interesting to observe that even without the inclusion of a 3NF,
the predictions by the Hannover-Lisbon approach for cross sections significantly differ from those by the Bochum-Cracow 
calculation. This points to a large sensitivity of the treatment of MEC. We, therefore, expect that the dominant part
of the observed deficiency of the theoretical predictions when compared to the experimental data stems from an incomplete 
modeling of MECs. For the analyzing powers, this deficiency is less pronounced and it seems that the addition 3NF within the Hannover-Lisbon approach brings the predictions closer to the data.
\section{Conclusions}
\label{Conc}
Differential cross sections of the deuteron-proton radiative capture at 55, 66.5 and 90~MeV/nucleon were measured 
with a coincidence setup. These cross sections were obtained almost background free due to the fact that a pure 
hydrogen target was used in the experiment and that both outgoing particles were measured in time coincidence. 
The data are compared with a few theoretical predictions constructed based on modern two- and three- nucleon 
potentials and taking MECs into account. The magnitude of the cross sections and analyzing powers from previous publication~\cite{ali} agree more with the model which 
incorporates the three-nucleon forces through the implementation of the $\Delta$ resonance in a coupled-channel 
approach. However, the widths of the angular distributions of the cross section data seem to behave differently from the predictions. 
The disagreement is the largest for 66.5 MeV/nucleon. These discrepancies cannot be explained by taking into account 
the systematic uncertainties. Both calculations predict that the effect of the three-nucleon forces 
at these energies are very small. We, therefore, expect that the discrepancies between data and the predictions of 
both models are strongly related to the implementation and the ingredients of MECs. The large differences among 
both models support this conclusion as well. 
\begin{acknowledgement}
The authors acknowledge the work by the cyclotron and ion-source groups at KVI for delivering the high-quality beam 
and technical supports used in this experiment. We also thank the theoretical groups at Cracow-Bochum and 
Hannover-Lisbon for their valuable numerical calculations. The authors thank GSI for the loan of the Plastic Ball. This work is part of the research program of the 
``Stichting voor Fundamenteel Onderzoek der Materie'' (FOM) with financial support from the 
``Nederlandse Organisatie voor Wetenschappelijk Onderzoek" (NWO).  Furthermore, the present work has been 
performed with financial support from the University of Groningen (RuG), the GSI Helmholtzzentrum f\"ur Schwerionenforschung 
GmbH, Darmstadt, and the Office of Research Vice-Chancellor of the University of Sistan and Baluchestan (USB) under Grant No. 892-2-166. L. Joulaeizadeh is thanked for her help in making the first figure of the paper. 
\end{acknowledgement} 
%
%

\end{document}